\documentstyle[preprint,aps]{revtex}
\begin{document}
\draft
\preprint{UM-P-94/118, RCHEP-33}
\title{From $m_{d}=m_{e}$ to Realistic Mass Relations in Quark-Lepton
Symmetric Models.}
\author{D. S. Shaw and R. R. Volkas\footnote{Address until Dec. 1994:
Department of Applied Mathematics and Theoretical Physics, University of
Cambridge, Cambridge, United Kingdom, CB3 9EW.} }
\address{Research Centre for High Energy Physics, School of Physics,\\
University of Melbourne, Parkville 3052, Australia.}
\maketitle

\begin{abstract}

In recent years a new potential symmetry of fundamental particle
physics has been investigated --- discrete
quark-lepton symmetry. When this symmetry is
implemented, however, it often leads to either of the unrealistic predictions
$m_{u}=m_{e}$ or $m_{d}=m_{e}$. This paper considers two possible ways models
based on $m_{d}=m_{e}$ can be made realistic.

\end{abstract}

\section{Introduction}

There are many outstanding questions left unanswered by the standard model (SM)
of particle physics. Various approaches have been used to address these
problems,
but one of the most common is the introduction of new symmetries to the
Lagrangian, often gauge symmetries, and usually only unbroken above some
threshold energy. Indeed, it is worth searching for possible new symmetries in
order to see what simplifications or interesting new relations they may
provide.
One such new symmetry uncovered recently is quark-lepton
symmetry~\cite{orig,LRvers}. This is a $Z_{2}$
discrete symmetry which seeks to put quarks and leptons on an equal footing. In
order to do this, a new gauge group, the $SU(3)_{l}$ group of leptonic colour,
needs to be introduced to parallel the standard quark colour group.

The introduction of this new symmetry provides a pleasing simplification of the
particle spectrum, but can also entail an unrealistic equality between the
masses
of quarks and leptons. It is certainly possible to circumvent this unwanted
equality while retaining the symmetry which initially lead to it, however a
more intriguing possibility is to treat the mass equalities as a correct and
promising starting point from which to obtain an explanation for at least some
aspects of the observed mass spectrum. In this paper, we shall examine a couple
of attempts to achieve just this aim, in the context of a left-right symmetric
generalisation of the basic theory.

In Sec.\ref{review}, the basic details of the quark-lepton-symmetric theory are
reviewed, along with the left-right symmetric generalisation.
Sec.\ref{compbroke} looks at the idea of completely broken leptonic
colour in the framework of the left-right symmetric version of the theory. This
is an extension of work done by one of the authors in a previous paper
\cite{ray}.
In Sec.\ref{seesawlike} a variant on the seesaw concept is implemented in order
to
obtain a suitable adjustment of the mass relations. Sec.\ref{conc} summarises
and concludes the paper.

\section{Review of Quark-Lepton Symmetry}
\label{review}

In the basic implementation of quark-lepton symmetry, the SM gauge group is
extended to $G_{ql} = SU(3)_{l}\otimes SU(3)_{q}\otimes SU(2)_{L}\otimes
U(1)_{X}$
which is unbroken at energies of order 1 TeV or more. The particle fields are:
\begin{equation} \begin{array}{cc}
Q_{L} \sim (3,1,2)(\frac{1}{3}), & \\
u_{R} \sim (3,1,1)(\frac{4}{3}), & d_{R} \sim (3,1,1)(-\frac{2}{3}), \\
F_{L} \sim (1,3,2)(-\frac{1}{3}), & \\
E_{R} \sim (1,3,1)(-\frac{4}{3}), & N_{R} \sim (1,3,1)(\frac{2}{3}), \\
\end{array} \end{equation}
and the Higgs fields are:
\begin{equation} \begin{array}{c}
\phi \sim (1,1,2)(1), \\
\begin{array}{cc}
\chi_{1} \sim (3,1,1)(\frac{2}{3}), & \chi_{2} \sim (1,3,1)(-\frac{2}{3}). \\
\end{array} \\ \end{array} \end{equation}

The first stage of symmetry breaking, resulting from the generation of a vacuum
expectation value (vev) $\langle\chi\rangle$ for $\chi_{1}$, will reduce the
gauge group factor $SU(3)_{l}\otimes U(1)_{X}$ to $SU(2)'\otimes U(1)_{Y}$,
where
\begin{equation}
Y = X + T_{8}/3,
\end{equation}
where $T_{8}$ is the diagonal generator diag(-2,1,1) of leptonic colour.
At this point the leptonic-colour triplets $F_{L}$, $E_{R}$ and $N_{R}$ will
break up into the familiar lepton fields, and into $SU(2)'$ doublet
``liptons'',
which will form into $SU(2)'$-singlet composite particles (since $SU(2)'$
should
be confining) that will decouple from the theory at energies below $SU(3)_{l}$
unification. After this, the theory follows the same pattern as the SM, with
$\phi$ developing a vev and breaking $SU(2)_{L}\otimes U(1)_{Y}$ down to
$U(1)_{\rm em}$.

However, because of the presence of the extra discrete quark-lepton symmetry,
\begin{equation} \begin{array}{cccc}
Q_{L} \leftrightarrow F_{L}, & u_{R} \leftrightarrow E_{R}, & d_{R}
\leftrightarrow N_{R}, & \phi \leftrightarrow \phi^{C}, \\
G_{q}^{\mu} \leftrightarrow G_{l}^{\mu}, & W^{\mu} \leftrightarrow W^{\mu}, &
C^{\mu} \leftrightarrow -C^{\mu}, & \\
\label{symmswaps} \end{array} \end{equation}
the Yukawa mass terms in the Lagrangian will be constrained to be of the form
\begin{equation}
{\cal L}_{\rm Yukawa} = \lambda_{1}(\overline{Q_{L}}u_{R}\phi^{C} +
\overline{F_{L}}E_{R}\phi) + \lambda_{2}(\overline{Q_{L}}d_{R}\phi +
\overline{F_{L}}N_{R}\phi^{C}) + H.c.,
\end{equation}
where $\phi^{C} = i\tau_{2}\phi$, thus resulting in tree level predictions for
the quark and lepton masses:
\begin{equation} \begin{array}{cc}
m_{u} = m_{e}, & m_{d} = m_{\nu}. \\
\end{array} \label{massrels1} \end{equation}
The seesaw mechanism~\cite{seesaw} can be invoked to cure the second of these
relations, by --- for example --- the introduction of new Higgs fields
\begin{equation} \begin{array}{cc}
\Delta_{1} \sim (\overline{6},1,1)(\frac{4}{3}), &
\Delta_{2} \sim (1,\overline{6},1)(-\frac{4}{3}), \\
\end{array} \end{equation}
but the first remains, and is clearly not in good agreement with experiment.

There are two possible variants if left-right symmetry is simultaneously
imposed
on the theory~\cite{LRvers}. In the first variant essentially the same model
results, so that
no new information relating to the promise, and problem, of the relations of
Eq.~(\ref{massrels1}) is obtained. In the second variant, however, the
relations
are switched, generating the different (if still problematic) relations
\begin{equation} \begin{array}{cc}
m_{u} = m_{\nu}, & m_{d} = m_{e}. \\
\end{array} \label{massrels2} \end{equation}
Again, the seesaw mechanism can be invoked to cure one of these relations, but
not both. The gauge group for the left-right symmetric version of the theory is
$G_{qlLR} = SU(3)_{q}\otimes SU(3)_{l}\otimes SU(2)_{L}\otimes SU(2)_{R}\otimes
U(1)_{X'}$ and the particle fields are:
\begin{equation} \begin{array}{cc}
F_{L} \sim (3,1,2,1)(-\frac{1}{3}), & F_{R} \sim (3,1,1,2)(-\frac{1}{3}), \\
Q_{L} \sim (1,3,2,1)(\frac{1}{3}), & Q_{R} \sim (1,3,1,2)(\frac{1}{3}).
\label{fermfields} \end{array} \end{equation}
The above mass relations, Eq.~(\ref{massrels2}), then follow from the
Lagrangian
terms
\begin{equation}
{\cal L}_{qlLR} = \lambda_{1}(\overline{F_{L}}F_{R} +
\overline{Q_{L}}Q_{R})\Phi+
\lambda_{2}(\overline{F_{L}}F_{R} + \overline{Q_{L}}Q_{R})\Phi^{C} + H.c.,
\end{equation}
where $\Phi \sim (1,1,2,2)(0)$ is the Higgs field. By contrast with $\phi$ in
Eq.~(\ref{symmswaps}), the electroweak Higgs $\Phi$ is taken to transform into
itself under the discrete quark-lepton symmetry.

\section{Completely Broken Leptonic Colour in the qlLR Model}
\label{compbroke}

Clearly, the relations Eq.~(\ref{massrels1}) and Eq.~(\ref{massrels2}) above
must
be either circumvented or adjusted in some way. Previous work has looked at
both such possibilities. Here, we shall extend the work done in \cite{ray} to
the
case of left-right symmetric leptonic colour. In the original formulation of
quark-lepton symmetry, the leptonic colour group was only partially broken down
to an $SU(2)'$ subgroup [along with a $U(1)$ factor]. In the approach of
\cite{ray}, the leptonic colour group is
completely broken, leaving no confining subgroup. Obviously, this implies the
existence of free ``liptons'' of comparable or higher
mass to the standard quarks and
leptons. For an appropriate choice of the formula for the hypercharge, however,
this apparent failing can instead become quite beneficial. The liptons will
then
possess the same quantum numbers under the SM group as the regular leptons, so
that mass mixing will occur. It is possible to arrange the mass matrices so
that they naturally lead to a lepton mass eigenstate suppressed with respect to
its ql-symmetric quark partner, along with very large values for the other
eigenstates, in a manner not dissimilar to the universal seesaw
mechanism~\cite{univss}. In
\cite{ray}, charged lepton masses had to be suppressed relative to up quark
masses. In this paper, we will instead look at suppressing them relative to
down quark masses. This may be a more natural approach for the second and third
generations, since the $s$/$\mu$ and $b$/$\tau$ mass splittings are much
smaller than $c$/$\mu$ and $t$/$\tau$, respectively.

The fermion fields remain unchanged from those shown in Eq.~(\ref{fermfields}).
The Higgs content of the model is
\begin{equation} \begin{array}{c}
\Phi \sim (1,1,2,2)(0), \\ \begin{array}{cc}
\chi_{1} \sim (3,1,1,1)(\frac{2}{3}), & \chi_{2} \sim (1,3,1,1)(-\frac{2}{3}),
\\
\Delta_{1L} \sim (\overline{6},1,3,1)(\frac{2}{3}), & \Delta_{1R} \sim
(\overline{6},1,1,3)(\frac{2}{3}), \\
\Delta_{2L} \sim (1,\overline{6},3,1)(-\frac{2}{3}), & \Delta_{2R} \sim
(1,\overline{6},1,3)(-\frac{2}{3}). \\ \end{array} \\
\end{array} \end{equation}
The Yukawa Lagrangian for this model is
\begin{equation} \begin{array}{lll}
{\cal L} & = & \lambda_{1}(\overline{Q_{L}}Q_{R} + \overline{F_{L}}F_{R})\Phi +
\lambda_{2}(\overline{Q_{L}}Q_{R} +  \overline{F_{L}}F_{R})\Phi^{C} + \\
& & \lambda_{3}[\overline{(F_{L})^{C}}F_{L}\chi_{1} +
\overline{(Q_{L})^{C}}Q_{L}\chi_{2} +
\overline{(F_{R})^{C}}F_{R}\chi_{1} + \overline{(Q_{R})^{C}}Q_{R}\chi_{2}] + \\
& & \lambda_{4}[\overline{(F_{L})^{C}}F_{L}\Delta_{1L} +
\overline{(F_{R})^{C}}F_{R}\Delta_{1R} + \overline{(Q_{L})^{C}}Q_{L}\Delta_{2L}
+
\overline{(Q_{R})^{C}}Q_{R}\Delta_{2R}] + H.c.
\label{lang1} \end{array} \end{equation}
The full symmetry of the model is then broken down via various vevs. These vevs
are
\begin{equation}
\langle\Phi\rangle = \left(
\begin{array}{cc}
0 & f \\
f & 0 \\
\end{array}\right),
\end{equation}
\begin{equation}
\langle\chi_{1}\rangle = \left(
\begin{array}{c}
\omega \\ 0 \\ 0 \\
\end{array}\right),
\end{equation}
and
\begin{eqnarray}
\langle\Delta_{1R}\rangle & = & \left(
\begin{array}{ccc}
v_{1} & 0 & v_{2} \\
0 & 0 & 0 \\
v_{2} & 0 & v_{3} \\
\end{array}\right), t_{R3} = -1 \nonumber \\
 & & \left( \begin{array}{ccc}
0 & v_{4} & 0 \\
v_{4} & 0 & 0 \\
0 & 0 & 0 \\
\end{array}\right), t_{R3} = 0 \nonumber \\
& & \left( \begin{array}{ccc}
0 & 0 & 0 \\
0 & v_{5} & 0 \\
0 & 0 & 0 \\
\end{array}\right), t_{R3} = +1, \nonumber \\
\end{eqnarray}
where $t_{R3}$ is the diagonal generator of $SU(2)_{R}$.
The vevs $\langle\chi_{1}\rangle$ and $\langle\Delta_{1R}\rangle$ will break
the
gauge group down from $G_{qlLR}$ to $G_{SM}$, breaking the left-right symmetry
and completely breaking the leptonic colour group. The hypercharge $Y$ is
chosen
to have the form
\begin{equation}
Y = X + T_{8}/3 + T_{3} + t_{R3}/2,
\end{equation}
where $T_{3}$ is the diagonal generator diag(0,1,-1) of $SU(3)_{l}$.
The vev $\langle\Phi\rangle$ will then break the SM group down in the usual
manner, so that the electromagnetic charge $Q$ is given by
\begin{equation}
Q = Y/2 + t_{L3},
\end{equation}
where $t_{L3}$ is the diagonal generator of $SU(2)_{L}$.
If we now look at the components of the leptonic fields $F_{L}$ and $F_{R}$, we
find
\begin{equation}
Y(F_{L}) = Y\left(\begin{array}{c}
l_{1L} \\ (l_{2R})^{C} \\ l_{3L} \\ \end{array}\right) = \left(\begin{array}{c}
-1 \\ +1 \\ -1 \\ \end{array}\right) \sim \left(\begin{array}{c}
(\nu_{1L}, e_{1L}) \\ ((e_{2R})^{C}, (\nu_{2R})^{C}) \\ (\nu_{3L}, e_{3L}) \\
\end{array} \right),
\end{equation}
\begin{equation}
Y(F_{R}) = Y\left(\begin{array}{c}
l_{1R} \\ (l_{2L})^{C} \\ l_{3R} \\ \end{array}\right) = \left(\begin{array}{c}
-1 \\ +1 \\ -1 \\ \end{array}\right) \sim \left(\begin{array}{c}
(\nu_{1R}, e_{1R}) \\ ((e_{2L})^{C}, (\nu_{2L})^{C}) \\ (\nu_{3R}, e_{3R}) \\
\end{array} \right).
\end{equation}

If we now input the vevs $\langle\Phi\rangle, \langle\chi_{1}\rangle$ and
$\langle\Delta_{1R}\rangle$ into the Lagrangian, Eq.~(\ref{lang1}) above, we
get,
for the charged lepton sector,
\begin{equation} \label{ch-lep-masses}
(\overline{e_{1L}},\overline{e_{2L}},\overline{e_{3L}}) = \left(
\begin{array}{ccc}
m_{d} & 0 & 0 \\
M_{4} & m_{u} & M_{\omega} \\
0 & M_{\omega} & m_{d} \\
\end{array} \right) \left( \begin{array}{c}
e_{1R} \\ e_{2R} \\ e_{3R} \\ \end{array}\right),
\end{equation}
where $M_{4} = \lambda_{4}v_{4}$ and $M_{\omega} = \lambda_{3}\omega$.
In general, each of the above entries is a $3 \times 3$ matrix in generation
space. The three generational case is very complicated, but an idea of the
effects on the mass eigenvalues can be seen by considering the matrix formed
by taking into consideration only one generation. We find then, taking into
account the fact that $v_{4},\omega \gg m_{u}, m_{d}$, this matrix
diagonalises, to lowest order, to
\begin{equation}
{\rm diag}(M_{\omega}, \sqrt{M_{\omega}^{2} + M_{4}^{2}}, m_{e}),
\end{equation}
where
\begin{equation}
m_{e} \approx m_{d}\cos\phi \leq m_{d},
\end{equation}
and
\begin{equation}
\cos\phi = \frac{M_{\omega}}{\sqrt{M_{\omega}^{2} + M_{4}^{2}}}.
\end{equation}

        Thus, within each generation, the charged lepton will be left with a
suppressed mass with respect to its down quark partner. Furthermore, the size
of this suppression is dependant on $M_{4}$ and $M_{\omega}$, which may vary
from generation to generation, thus allowing for the correct level of
suppression within each generation. This will be further complicated by any
intergenerational mixing.

        Next we turn to the neutrino fields. Here there are three matrices to
look at. First, there is the Dirac-mass matrix between left- and right-handed
neutrino fields:

\begin{equation}
(\overline{\nu_{1L}},\overline{\nu_{2L}},\overline{\nu_{3L}}) =
\left( \begin{array}{ccc}
m_{u} & 0 & 0 \\
M_{4} & m_{d} & M_{\omega} \\
0 & M_{\omega} & m_{u} \\ \end{array} \right) \left( \begin{array}{c}
\nu_{1R} \\ \nu_{2R} \\ \nu_{3R} \\ \end{array} \right).
\end{equation}

This matrix produces similar results to the charged lepton matrix,
Eq.~(\ref{ch-lep-masses}),
generating two heavy neutrinos and one light neutrino, with a mass close to
the corresponding up-quark mass, but suppressed by $\cos\phi$. The second
neutrino mass matrix couples the right-handed neutrino fields to each other,

\begin{equation}
(\overline{(\nu_{1R})^{C}},\overline{(\nu_{2R})^{C}},\overline{(\nu_{3R})^{C}})
=
\left( \begin{array}{ccc}
\lambda_{4}v_{1} & 0 & \lambda_{4}v_{2} \\
0 & 0 & 0 \\
\lambda_{4}v_{2} & 0 & \lambda_{4}v_{3} \\ \end{array} \right)
\left( \begin{array}{c}
\nu_{1R} \\ \nu_{2R} \\ \nu_{3R} \\ \end{array} \right),
\end{equation}
while the final matrix couples left-hand neutrino fields:
\begin{equation}
(\overline{\nu_{1L}},\overline{\nu_{2L}},\overline{\nu_{3L}}) =
\left( \begin{array}{ccc}
0 & 0 & 0 \\
0 & \lambda_{4}v_{5} & 0 \\
0 & 0 & 0 \\ \end{array} \right)
\left( \begin{array}{c}
(\nu_{1L})^{C} \\ (\nu_{2L})^{C} \\ (\nu_{3L})^{C} \\ \end{array} \right).
\end{equation}

These last two matrices provide heavy Majorana masses (with the mass scale set
by the size of the vevs $v_{1}$, $v_{2}$, $v_{3}$, $v_{5}$) that will ensure
all the physical neutrino mass eigenstates are heavy via the standard seesaw
mechanism.

        Thus far, we have largely ignored the effects of inter-generational
mixing. It may be possible, however, that the very mass splitting we have
achieved here between the down-quark and charged-lepton masses might follow
from radiative corrections due to such inter-generational mixing. This was
discussed by Foot and Lew in ref.~\cite{radiat}. Here, the mass
splitting between the tauon and the bottom quark was addressed, with a view to
showing that even such a large gap could be bridged by radiative corrections
involving the much heavier top quark. The story becomes much more complicated
for the lighter generations, but the same principles would apply. This would
hopefully resolve the mass disparity satisfactorily without the need to invoke
the more complicated symmetry breaking scenario used in this paper or to break
the leptonic colour group down completely.

        In their paper, Foot and Lew did indeed show that radiative corrections
could be of sufficient size to explain the mass difference between the tauon
and the bottom quark. The precise size of these corrections, however, is
dependant on unknown parameters, and could vary over a large range of values.
The sign of the
corrections, also, is not easily fixed, so that it is possible that, far from
helping the situation, the radiative corrections
exacerbate the problem in some regions of parameter space. By
contrast, the method described in this paper, at least for the case of no
intergenerational mixing, clearly results in a suppression of the charged
lepton
masses.

\section{Seesaw Approach}
\label{seesawlike}

In this second approach, we look more directly towards a seesaw-like method of
suppressing the lepton masses. Again, the left-right-symmetric version of the
basic theory is used. The idea is to invoke a seesaw-like
mechanism to do for the electron what the standard seesaw idea does for the
neutrino. To do this, singlet fermion fields are introduced which will pick up
mass contributions from mixing with each other and with the regular leptonic
fields, thus generating a mass matrix with the traditional seesaw form.
Specifically, we have the fermion content shown in Eq.~(\ref{fermfields}), in
three copies - one for each generation - and in addition, one copy of the
following new fermion fields:
\begin{equation} \begin{array}{cc}
\epsilon_{L} \sim (1,1,1,1)(-2), & \epsilon_{R} \sim (1,1,1,1)(-2) \\
\end{array} \end{equation}

The Higgs content for this idea is as follows:
\begin{equation} \begin{array}{c}
\Phi \sim (1,1,2,2)(0), \\ \begin{array}{cc}
\chi_{1} \sim (3,1,1,1)(\frac{2}{3}), & \chi_{2} \sim (1,3,1,1)(-\frac{2}{3}),
\\
\Delta_{1L} \sim (\overline{6},1,3,1)(\frac{2}{3}), &
\Delta_{1R} \sim (\overline{6},1,1,3)(\frac{2}{3}), \\
\Delta_{2L} \sim (1,\overline{6},3,1)(-\frac{2}{3}), &
\Delta_{2R} \sim (1,\overline{6},1,3)(-\frac{2}{3}), \\
\eta_{1L} \sim (\overline{3},1,2,1)(-\frac{5}{3}), &
\eta_{1R} \sim (\overline{3},1,1,2)(-\frac{5}{3}), \\
\eta_{2L} \sim (1,\overline{3},2,1)(\frac{5}{3}), &
\eta_{2R} \sim (1,\overline{3},1,2)(\frac{5}{3}). \\ \end{array} \\
\end{array} \end{equation}

The Higgs and particle fields will couple according to the Yukawa mass
Lagrangian
\begin{equation}
{\cal L}_{\rm Yuk} = {\cal L}_{\rm Dirac} + {\cal L}_{\rm liptons} +
{\cal L}_{\rm Majorana} + {\cal L}_{\rm seesaw},
\end{equation}
where
\begin{eqnarray}
{\cal L}_{\rm Dirac} = \lambda_{1}(\overline{F_{L}}F_{R} +
\overline{Q_{L}}Q_{R})\Phi + \lambda_{2}(\overline{F_{L}}F_{R} +
\overline{Q_{L}}Q_{R})\Phi^{C} + H.c., \\
{\cal L}_{\rm liptons} = \lambda_{3}[\overline{(F_{L})^{C}}F_{L}\chi_{1} +
\overline{(F_{R})^{C}}F_{R}\chi_{1} + \overline{(Q_{L})^{C}}Q_{L}\chi_{2} +
\overline{(Q_{R})^{C}}Q_{R}\chi_{2}] + H.c., \\
{\cal L}_{\rm Majorana} = \lambda_{4}[\overline{(F_{L})^{C}}F_{L}\Delta_{1L} +
\overline{(F_{R})^{C}}F_{R}\Delta_{1R} +
\overline{(Q_{L})^{C}}Q_{L}\Delta_{2L}+
\overline{(Q_{R})^{C}}Q_{R}\Delta_{2R}] + H.c., \\
{\cal L}_{\rm seesaw} = \lambda_{5}(\overline{F_{L}}\epsilon_{R}\eta_{1L} +
\overline{\epsilon_{L}}F_{R}\eta_{1R} +
\overline{Q_{L}}(\epsilon_{L})^{C}\eta_{2L} +
\overline{(\epsilon_{R})^{C}}Q_{R}\eta_{2R}) +
M_{\epsilon\epsilon}\overline{\epsilon_{L}}\epsilon_{R} + H.c., \nonumber \\
\end{eqnarray}
where $M_{\epsilon\epsilon}$ is a bare mass term [or alternatively,
$M_{\epsilon\epsilon}$ results from the vev of a Higgs field $\sigma \sim
(1,1,1,1)(0)$].

The gauge group is then broken down via non-zero vevs $\langle\Phi\rangle$,
$\langle\chi_{1}\rangle$, $\langle\Delta_{1R}\rangle$ and
$\langle\eta_{1R}\rangle$ according to the hierarchy
\begin{equation} \begin{array}{cc}
SU(3)_{l}\otimes SU(3)_{q}\otimes SU(2)_{L}\otimes SU(2)_{R}\otimes U(1)_{X} &
\\
\downarrow & \langle\chi_{1}\rangle, \langle\Delta_{1R}\rangle,
\langle\eta_{1R}\rangle \\
SU(3)_{q}\otimes SU(2)_{L}\otimes U(1)_{Y} & \\
\downarrow & \langle\Phi\rangle \\
SU(3)_{q}\otimes U(1)_{em}. & \\
\end{array} \end{equation}

 From ${\cal L}_{\rm Dirac}$ we get the basic quark-lepton mass equalities,
Eq.~(\ref{massrels2}). ${\cal L}_{\rm liptons}$ will lead to heavy masses for
the liptons.
${\cal L}_{\rm Majorana}$ will provide the usual neutrino seesaw mechanism.
Finally,
${\cal L}_{\rm seesaw}$ will generate mass terms involving the
new fermion fields $\epsilon_{L}, \epsilon_{R}$. Combining all the mass terms
involving the $\epsilon$ and charged lepton fields, we obtain a matrix of the
form
\begin{equation}
\left( \begin{array}{cccc}
m_{d} & 0 & 0 & 0 \\
0 & m_{s} & 0 & 0 \\
0 & 0 & m_{b} & 0 \\
\lambda_{5}\langle\eta_{1R}\rangle & \lambda_{5}\langle\eta_{1R}\rangle &
\lambda_{5}\langle\eta_{1R}\rangle & M_{\epsilon\epsilon} \\
\end{array} \right).
\end{equation}
If we assume $\lambda_{5}\langle\eta_{1R}\rangle$ is of order 200 GeV, and that
$M_{\epsilon\epsilon}$ is of order 5 GeV, then
upon diagonalisation, we find three mass eigenvalues suppressed with respect to
the
input quark masses, and one very heavy eigenvalue.

A numerical analysis was performed to see if realistic levels of suppression
could be achieved. It was found that suitable results were easy to obtain, with
very accurate values obtainable if one allows for a slight generational
hierarchy in $\lambda_{5}$ (which is a $3 \times 3$ matrix in generation
space). For
example, for the following input values:
\begin{equation}
\left( \begin{array}{cccc}
0.01 & 0 & 0 & 0 \\
0 & 0.2 & 0 & 0 \\
0 & 0 & 5 & 0 \\
100 & 150 & 500 & 5 \\
\end{array} \right).
\end{equation}
we obtain, upon diagonalisation, the eigenvalues
\begin{equation} \begin{array}{cccc}
m_{e} = 0.5 MeV, & m_{\mu} = 110 MeV, & m_{\tau} = 1.7 GeV, & M = 530 GeV, \\
\end{array} \end{equation}
which is clearly a good match with experiment.

\section{Conclusion}
\label{conc}

Although it provides a pleasing simplification of the fermion fields, a
discrete
quark-lepton symmetry can introduce an unrealistic mass relation between the
charged leptons and either the up-quark fields Eq.~(\ref{massrels1}) or the
down-quark fields Eq.~(\ref{massrels2}) (in the case of a left-right symmetric
version of the theory). Rather than simply circumventing this mass relation, in
this paper we have treated the relation as a sign post towards a greater
understanding of the fermion mass spectrum.

After summarising the basic framework of quark-lepton symmetry, we discussed
two ideas by which the mass relations Eq.~(\ref{massrels2}) might lead to
explanations for the major trends in the masses of the fermions within each
family.

The first of these efforts was an extension of the method used in \cite{ray},
in which the leptonic colour group is completely broken down by the various
Higgs field vevs, to consider the case of the left-right symmetric version of
the theory. There it was shown that the method would successfully suppress the
masses of the charged fermion fields with respect to those of the down quark
fields in the absence of intergenerational mixing. In comparison, the use of
radiative corrections to split these masses
(investigated in \cite{radiat}), does not necessarily lift the degeneracy
in the right direction.

In the second proposal, the addition of one generation of vector-like charged
lepton fields was shown, via a see-saw like mechanism, to also suppress the
charged lepton masses with respect to their quark counterparts. A numerical
analysis showed that realistic levels of suppression are possible with this
method.

\section{Aknowledgements}
R. R. V. would like to thank Prof. J. C. Taylor and the high energy theory
group at DAMTP, Cambridge University for hospitality while part of this
work was done. R. R. V. is supported by grants from the Australian Research
Council and the University of Melbourne. D. S. S. is supported by a research
award from the Department of Employment, Education and Training.

\end{document}